\documentclass[10pt,twoside,floatfix,twocolumn,final,superscriptaddress,showpacs,aps,prl]{revtex4-1}

\usepackage{graphicx,bm,amsmath,amssymb,color}
\usepackage[utf8]{inputenc}
\usepackage{soul}
\usepackage{times}

\newcommand{\overbar}[1]{\overline{#1}}

\begin{document}
\title{Spin Andreev-like Reflection in Metal-Mott Insulator Heterostructures}

\author{K. A. Al-Hassanieh}
\thanks{These authors contributed equally to this work.}
\affiliation{Center for Nanophase Materials Sciences, Oak Ridge National Laboratory, Oak Ridge, Tennessee 37831, USA}
\author{Juli\'an Rinc\'on}
\thanks{These authors contributed equally to this work.}
\affiliation{Center for Nanophase Materials Sciences, Oak Ridge National Laboratory, Oak Ridge, Tennessee 37831, USA}
\affiliation{Perimeter Institute for Theoretical Physics, Waterloo, Ontario N2L 2Y5, Canada}

\author{G. Alvarez}
\affiliation{Center for Nanophase Materials Sciences, Oak Ridge National Laboratory, Oak Ridge, Tennessee 37831, USA}
\affiliation{Computer Science \& Mathematics Division, Oak Ridge National Laboratory, Oak Ridge, Tennessee 37831, USA}

\author{E. Dagotto}
\affiliation{Materials Science and Technology Division, Oak Ridge National Laboratory, Oak Ridge, Tennessee 37831, USA}
\affiliation{Department of Physics and Astronomy, The University of Tennessee, Knoxville, Tennessee 37996, USA}

\date{\today}

\begin{abstract}
Using the time-dependent density-matrix renormalization group (tDMRG), 
we study the time evolution of electron wave packets in one-dimensional (1D)
metal-superconductor heterostructures. The results show 
Andreev reflection at the interface, as expected. By combining 
these results with the well-known single-spin-species 
electron-hole transformation in the Hubbard model, we predict an analogous
spin Andreev reflection in metal-Mott insulator heterostructures. 
This effect is numerically confirmed using 1D tDMRG, but it 
is expected to be present also in higher dimensions, as well as in more general
Hamiltonians. We present an intuitive picture of the spin reflection, 
analogous to that of Andreev reflection at metal-superconductors interfaces.  
This allows us to discuss a novel antiferromagnetic proximity effect. Possible experimental realizations are discussed.
\end{abstract}

\pacs{71.10.Fd, 71.10.Pm, 74.20.-z, 74.45.+c}

\maketitle

{\it Introduction}.---Correlated electrons at the interface 
of two materials can exhibit a wide range of remarkable phenomena. 
Among the most interesting effects is the Andreev reflection (AR) 
at the normal metal-superconductor (N-SC) interface~\cite{Andreev}. 
As an electron is transmitted from the normal metal (N) into the superconductor (SC), it attracts 
a second electron from the metal to form a Cooper pair. 
The second paired electron leaves a hole that is reflected off the 
interface into the metal. In AR a charge current in 
the metal becomes a supercurrent as it enters the SC. 

Several variants of AR have been widely discussed 
in recent years. Specular AR has been predicted 
at graphene N-SC interfaces, where the reflection angle is 
inverted~\cite{Beenakker}. Proximity effect and AR in 
SC-ferromagnet heterostructures have been 
studied both theoretically~\cite{Buzdin, de Jong} and 
experimentally~\cite{Robinson, Khaire, Visani}. 
Spin-dependent $Q$-reflection was predicted in normal 
metal-itinerant antiferromagnet (N-AFM) interfaces~\cite{Bobkova}, 
where a $\pi$ phase shift is seen between the reflected spin-up 
and spin-down electrons. In N-SC-N systems, nonlocal or crossed
AR (CAR) involves the transmission of an electron from a metal 
to the SC and the creation of a hole in the other metal. 
In transport experiments, CAR competes with elastic cotunneling  
as the dominant mechanism~\cite{Kleine}. Due to its nonlocal 
nature, CAR can generate nonlocal 
entanglement and correlations in carbon nanotubes~\cite{Herrmann} and 
topological superconductors~\cite{He}.  

In one-dimensional (1D) systems,  dominant superconducting 
fluctuations are characterized by the Luttinger liquid (LL) 
parameter $K_\rho > 1$, or $K > 1$ in the case of spinless 
fermions~\cite{Giamarchi2004}. AR was predicted for 1D spinless 
fermions with nearest-neighbor (NN) attraction $V < 0$ coupled 
to a noninteracting wire ($K = 1$). The reflection coefficient 
is given by $\gamma = (1 - K) / (1+ K)$. For $K > 1$, $\gamma < 0$, 
which corresponds to AR~\cite{Safi}. This formula can be generalized 
to the spin and charge sectors of electrons in 1D~\cite{Hou}. The spinless 
fermions results were later confirmed using time-dependent density matrix renormalization group (tDMRG) calculations 
that cast  the AR in the context 
of 1D cold atoms~\cite{Daley}. To our knowledge, a real-time study of AR 
in the case of spinful electrons has not been presented before. We are 
also unaware of any prediction of its spin analogue described in this work.  

In this Letter, we propose a novel spin Andreev-like reflection (SAR) 
in normal metal-Mott insulator (N-MI) heterostructures. An electron with 
a given spin projection undergoes a spin flip upon reflection, inducing 
a spin-1 excitation in the MI. We present 
an intuitive picture of the process, analogous to that of AR in N-SC.  
SAR is verified in a Hubbard chain using tDMRG. However, the effect is not 
restricted to 1D, and can be observed in higher dimensions as well. 
We discuss the differences of our work with previous efforts, and 
present possible experimental realizations of the new effect. 

The organization of our work is the following. First, we study 
the dynamics of a wave packet with both spin and charge 
colliding with a N-SC interface. The superconductor 
is modeled using the Hubbard Hamiltonian with onsite 
attraction $U < 0$. Away from half-filling, we observe the 
partial AR of the charge component. Due to the spin gap, 
the spin of the wave packet undergoes normal reflection. 
Then, by using a well-known single-spin-species electron-hole 
transformation, these results are translated into the novel 
SAR in N-MI systems. Finally, this translation is confirmed by 
calculating the time evolution of the wave packet colliding with 
the N-MI interface and clearly observing the spin Andreev reflection. 
We explore an AFM analogue of the SC proximity effect in the N-MI heterostructure. To close, we propose experimental realizations to test our predictions.

{\it Model}.---Using tDMRG~\cite{White,tdmrg,dmrgreview}, 
we study a Hubbard chain of $L_I$ sites connected to a non-interacting 
lead of $L_L$ sites. The Hamiltonian can be written as $H = H_I + H_T $, 
where $H_I$ represents the interactions in the Hubbard chain
\begin{equation}
H_I =  U\sum_{i = 1}^{L_I} \left(n_{i\uparrow} - \frac{1}{2}\right) \left(n_{i\downarrow} - \frac{1}{2}\right) + \mu \sum_{i = 1}^{L_I} n_i,
\end{equation}
and $H_T$ is the kinetic energy of the entire system,
\begin{equation}
H_T = -t_h\sum_{\sigma, i = 1}^{L_I + L_L -1} \left(c_{i\sigma}^\dagger c_{i+1\sigma}^{\;} + \textrm{H.c.}\right).
\end{equation}
The hopping integral $t_h$ is taken as the energy unit. The chemical potential $\mu$ of the Hubbard chain relative to the lead controls its charge density. The rest of the notation is standard. 
Note that $H$ has SU(2) symmetry in the spin sector and, for $\mu = 0$, the same symmetry in the charge sector. 
For details of the implementation of the time evolution of the wave packets see Refs.~\onlinecite{Al-Hassanieh13} and \onlinecite{suppl}.


{\it Results}.---For $U < 0$,  we study the half-filled case with $\mu = 0$, and $N_\uparrow = N_\downarrow = 30$, and a doped case with $N_\uparrow = N_\downarrow = 24$ 
and $\mu = -0.2$. At half-filling, the charge is perfectly transmitted and the spin is totally reflected due to the absence and presence of energy gaps, respectively. Therefore, we observe no evidence of AR. This can be understood by noticing that, for $U<0$, $K_\rho=1~(\gamma=0)$: the free-fermion limit~\cite{suppl}.

\begin{figure}
\centering
\includegraphics*[width=\columnwidth]{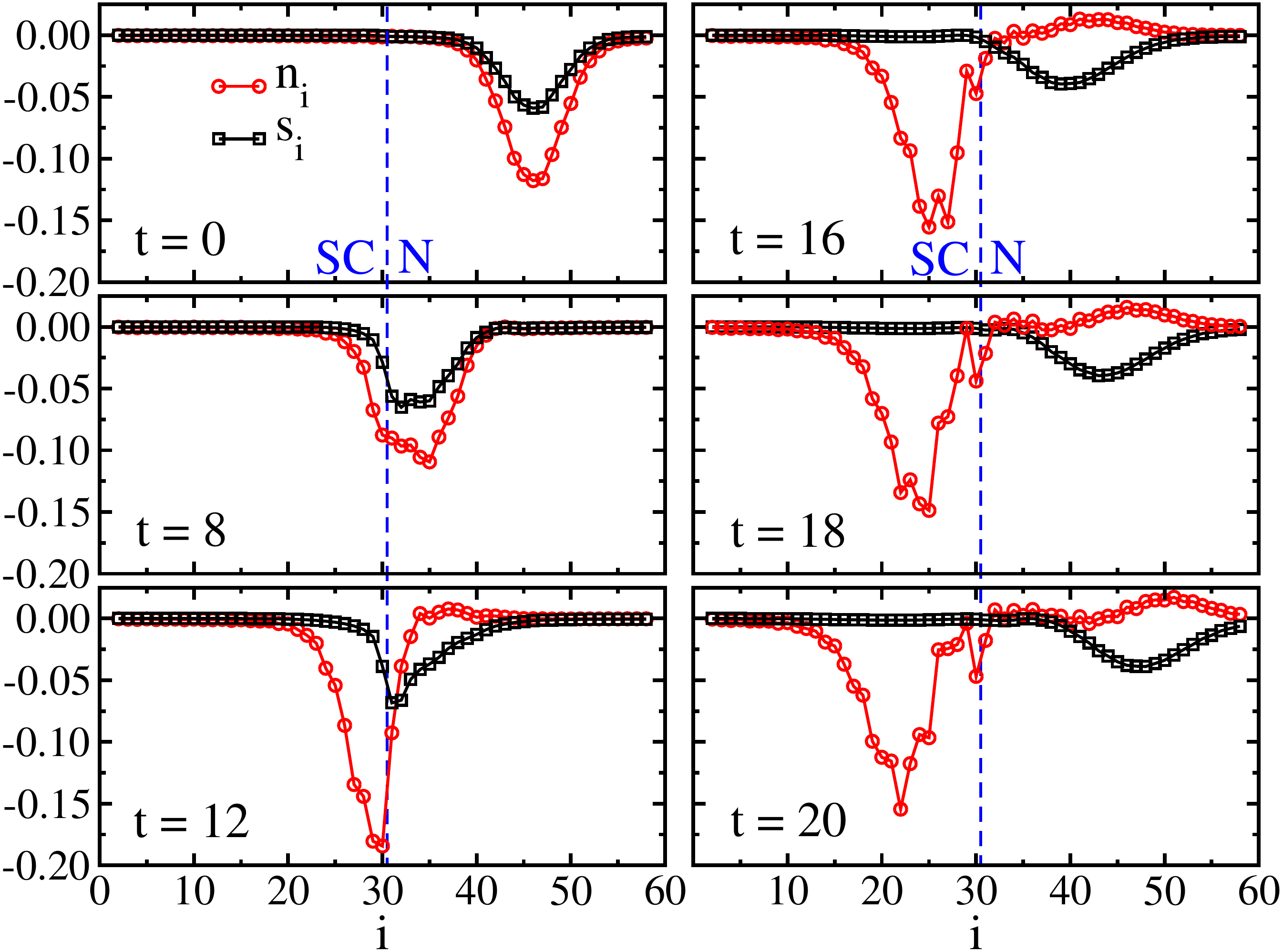}
\caption{Snapshots of the charge and spin propagation for $U = -4t_h$ in the doped case. The spin behavior is the same as in the half-filled case since the 
spin sector is still gapped, but the charge behavior is different. 
As it enters the interacting region, the charge of the wave packet increases, 
and consequently a small wave packet of opposite charge is formed in the lead, 
which is a clear signature of partial AR. This can be seen between $t = 12$ and $t = 20$. 
The reflected charge and spin move at the Fermi velocity of the lead. }
\label {Un4}
\end{figure}

The results for the doped case and  $U = -4t_h$ are shown in Fig.~\ref{Un4}. 
The behavior of the spin component is the same as 
in the half-filled case, since the spin sector is still gapped. 
The charge behavior is more interesting. After entering the 
SC region, the charge of the wave packet increases, 
while a small wave packet of opposite charge is formed at the lead. 
This is a clear signature of partial AR, i.e.~the reflection of 
an opposite charge at the interface. 
This reflected charge and spin move at the Fermi velocity of the lead. 
Note that in this case $K_\rho > 1$, leading to a negative reflection 
coefficient: this means that additional charge is attracted into the 
superconductor, or equivalently, opposite charge is reflected. 
For larger values of $|U|$, such as $U = -8t_h$, AR is not observed within our 
resolution and the charge undergoes normal reflection. This is consistent 
with previous studies in the regime where the SC pairing energy 
is comparable to the Fermi energy, as in ultracold fermionic mixtures, 
where it has been observed that the specular reflection of particles 
has a robust amplitude~\cite{BEC}. 

Consider now the main results of our publication. 
A well-known property of the Hubbard model with onsite interaction 
is the equivalence between the charge and spin sectors of the $U > 0$ 
and $U < 0$ cases. In other words, if the Hubbard model with $U < 0$ 
is studied, then the $U > 0$ properties can be deduced by merely 
exchanging charge and spin. More concretely, consider the 
electron-hole transformation on one spin species $\hat T$:
$c^{\dagger}_{i\uparrow} \to c^{\dagger}_{i\uparrow}$ and $c^{\dagger}_{i\downarrow} \to (-1)^ic_{i\downarrow}$, and consequently, 
$n_{i\uparrow} \to n_{i\uparrow}$ and $n_{i\downarrow} \to 1 - n_{i\downarrow}$. This transformation leaves $H_T$ invariant and maps $H_I$ 
into $\overbar H_I$ such that
\begin{equation}
\overbar H_I =  \overbar U\sum_{i = 1}^{L_I} \left(n_{i\uparrow} - \frac{1}{2}\right) \left(n_{i\downarrow} - \frac{1}{2}\right) + B \sum_{i = 1}^{L_I} S^z_i,
\end{equation}
where $\overbar U = -U$, $S^z_i = \frac{1}{2}(n_{i\uparrow} - n_{i\downarrow})$, and $B = 2\mu$ is the Zeeman field. In other words, $\hat T$ maps the attractive Hubbard model 
into its repulsive equivalent with the charge and spin sectors exchanged. 
In $\overbar H_I$, $B$ breaks the spin SU(2) symmetry, similarly as $\mu$ breaks 
the charge SU(2) symmetry in $H_I$.

\begin{figure}
\centering
\includegraphics*[width=\columnwidth]{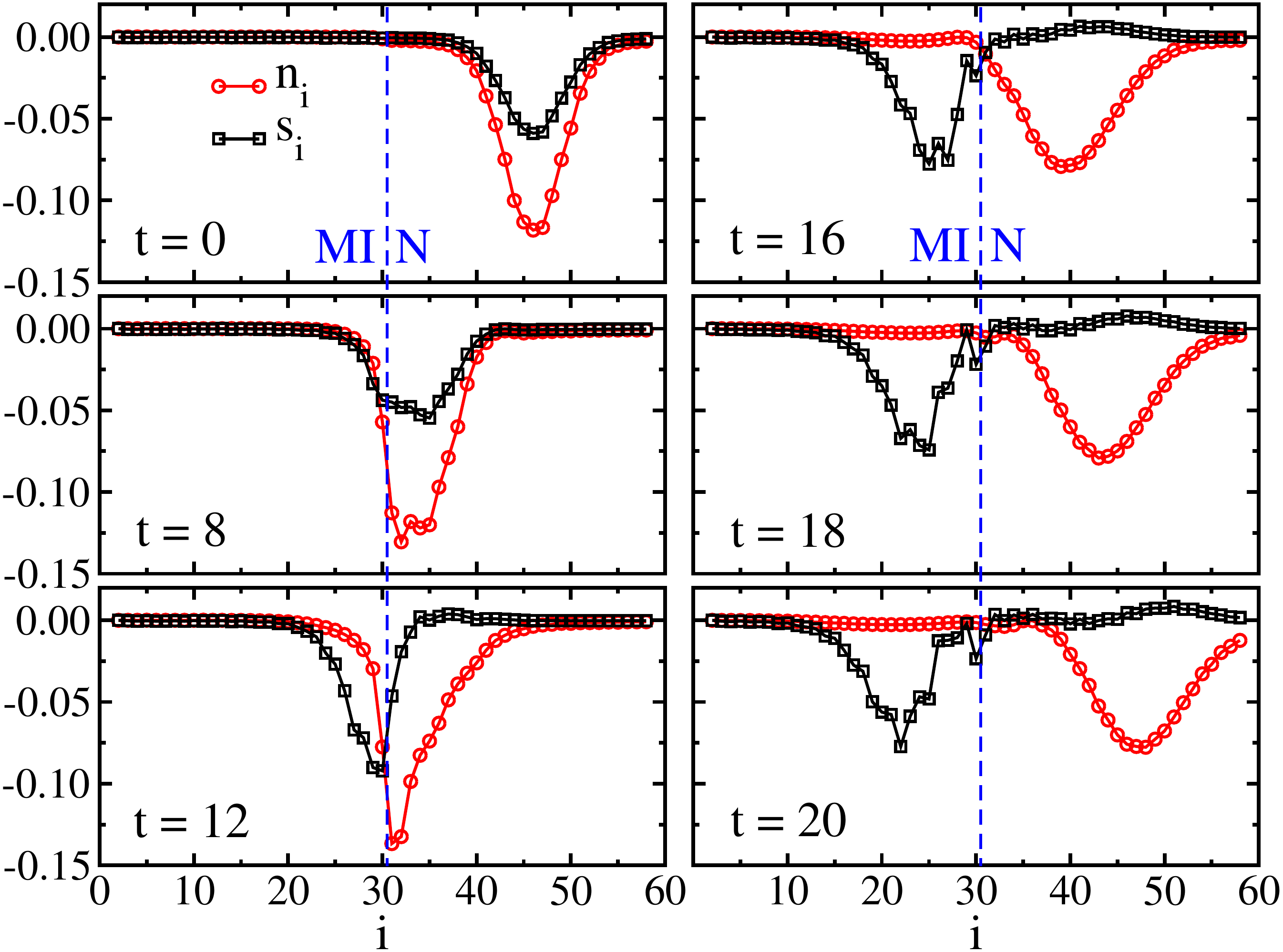}
\caption{Wave packet propagation for the repulsive Hubbard model with a 
finite magnetization (see text for parameters used). 
The wave packet spin is transmitted, whereas the charge 
is reflected due to the charge gap. SAR is observed. As the spin 
enters the MI, it expels an opposite spin back to the 
lead, i.e. an effective spin flip takes place. The spin flip here is partial, 
as in the charge AR case. This reflection 
of an opposite spin is the signature of the proposed SAR. }
\label {Up4}
\end{figure}

By applying this transformation to the results shown in Fig.~\ref{Un4}, 
or to the AR effect in general in any dimension, we can predict an interesting effect: 
an electron incident on a N-MI interface can undergo a spin flip upon reflection. 
To confirm this effect, the time evolution of wave packets was studied in 
the Hamiltonian $ \overbar H = H_T + \overbar H_I$. The parameters are 
set to those obtained from transforming $H$ using $\hat T$. That is, 
$\overbar U = 4t_h$, $B = -0.4$, $N_{\uparrow} = 24$ and  $N_{\downarrow} = 36$. 
The SC spin gap translates to the charge gap of the MI.  
The tDMRG results are shown in Fig.~\ref{Up4}. The spin behavior 
is identical to that of the charge shown in Fig.~\ref{Un4}, and vice versa. 
As the spin enters the MI, it expels an opposite spin back to the lead. 
In other words, an effective spin flip takes place (together with a normal
reflection without spin flip). The reflection of an opposite spin is the signature of the proposed SAR.  

\begin{figure}
\centering
\includegraphics*[width=\columnwidth]{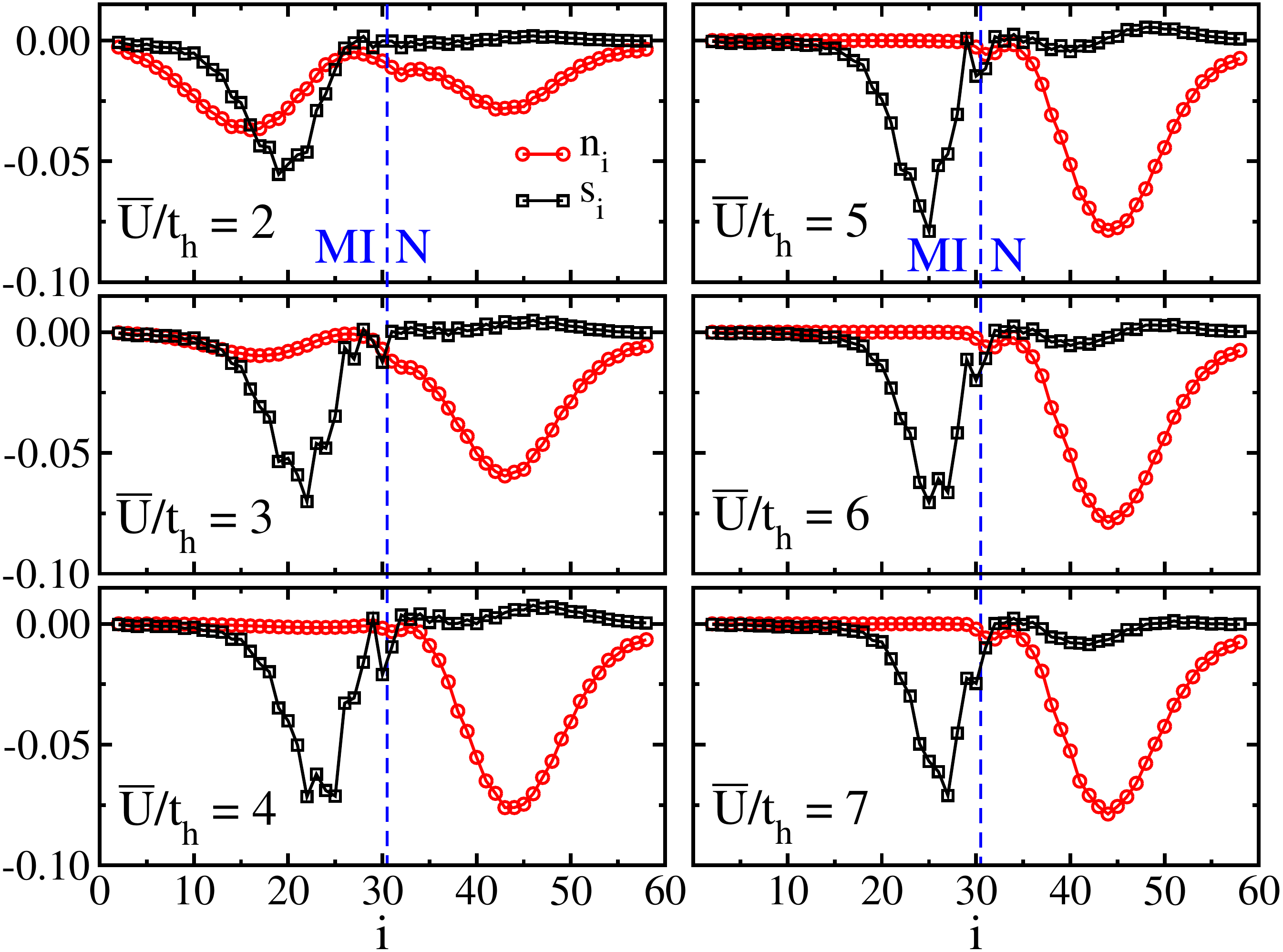}
\caption{ Charge and spin in the repulsive Hubbard model case for different values of $\overbar U$. The 
results are shown at $t = 18$.  SAR is maximal around $\overbar U = 4t_h$. For 
larger $\overbar U$, the SAR is suppressed and normal spin reflection becomes dominant. Note that for small $\overbar U$, part of the 
charge is transmitted due to the small charge gap and the finite energy of the wave packet. }
\label {U-2-7}
\end{figure}

The effect of electron interaction on SAR is shown in Fig.~\ref {U-2-7}. Snapshots of spin and charge propagation 
at $t = 18$ are shown for different values of $\overbar U$. In these calculations, we set $k_0 = k_{F\uparrow} - 2\sigma_k$, 
where $k_{F\uparrow}$ is the Fermi momentum of the spin-up electrons, and tune $B$ so that the magnetization in the system is 
uniform. SAR shows nonmonotonic behavior as a function of $\overbar U$. 
It is weak for small $\overbar U/t_h< 1$, where AFM fluctuations are unfavorable and free-fermion-like behavior is expected. Upon further increasing $\overbar U$, SAR reaches its optimal 
value at intermediate coupling around $\overbar U = 4t_h$. For $\overbar U/t_h\gg 1$, exchange and hence long-range order are suppressed as $J\sim t_h^2/\overbar U$. Also, normal spin reflection is observed in 
addition to SAR, and as $\overbar U$ increases, the normal reflection becomes dominant. This resulting trend is the AFM analogue of the BCS-BEC crossover observed in unitary Fermi gases~\cite{notebec,Randeria10}.

\begin{figure}
\centering
\includegraphics*[width=\columnwidth]{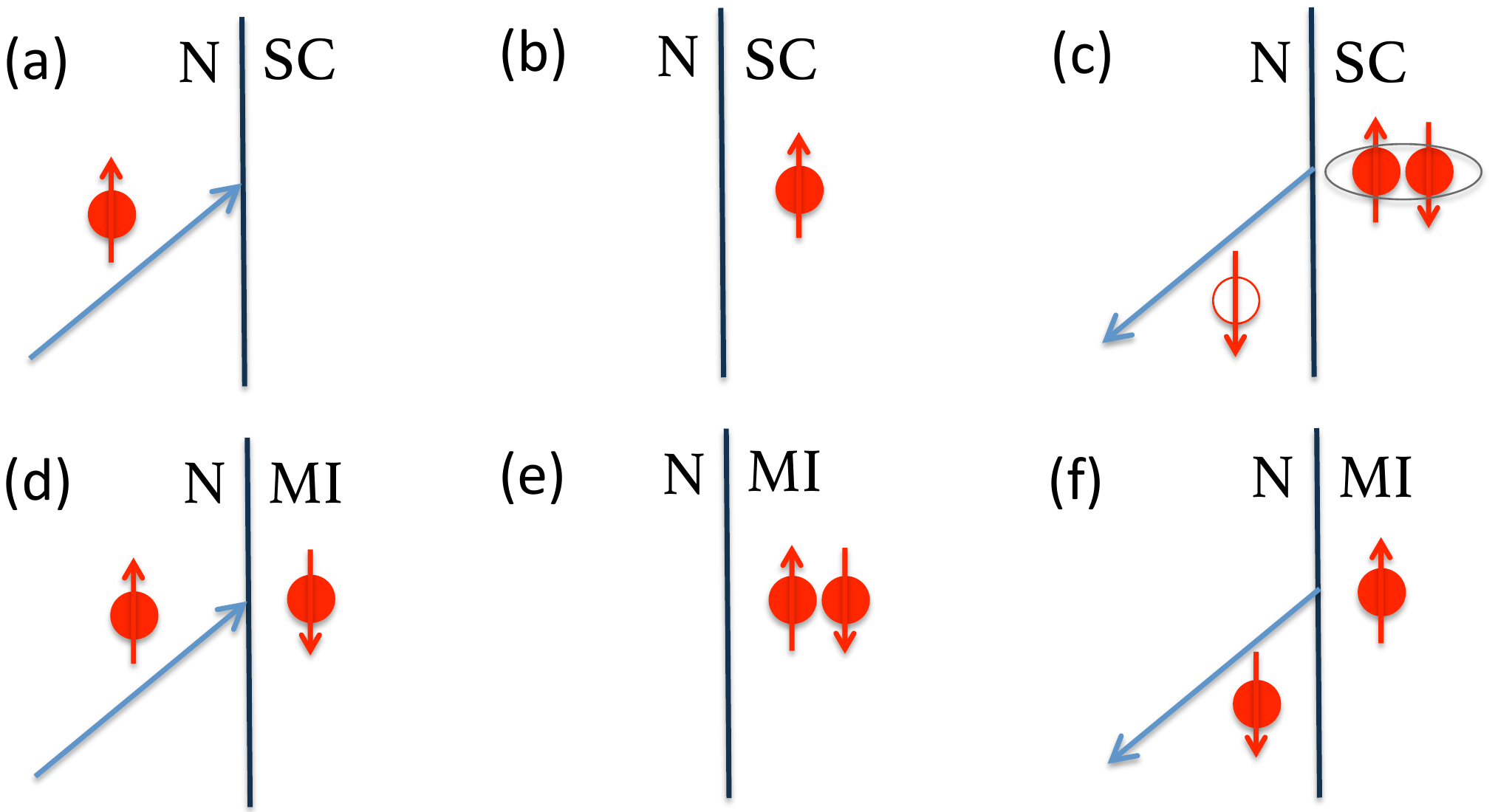}
\includegraphics*[width=.7\columnwidth]{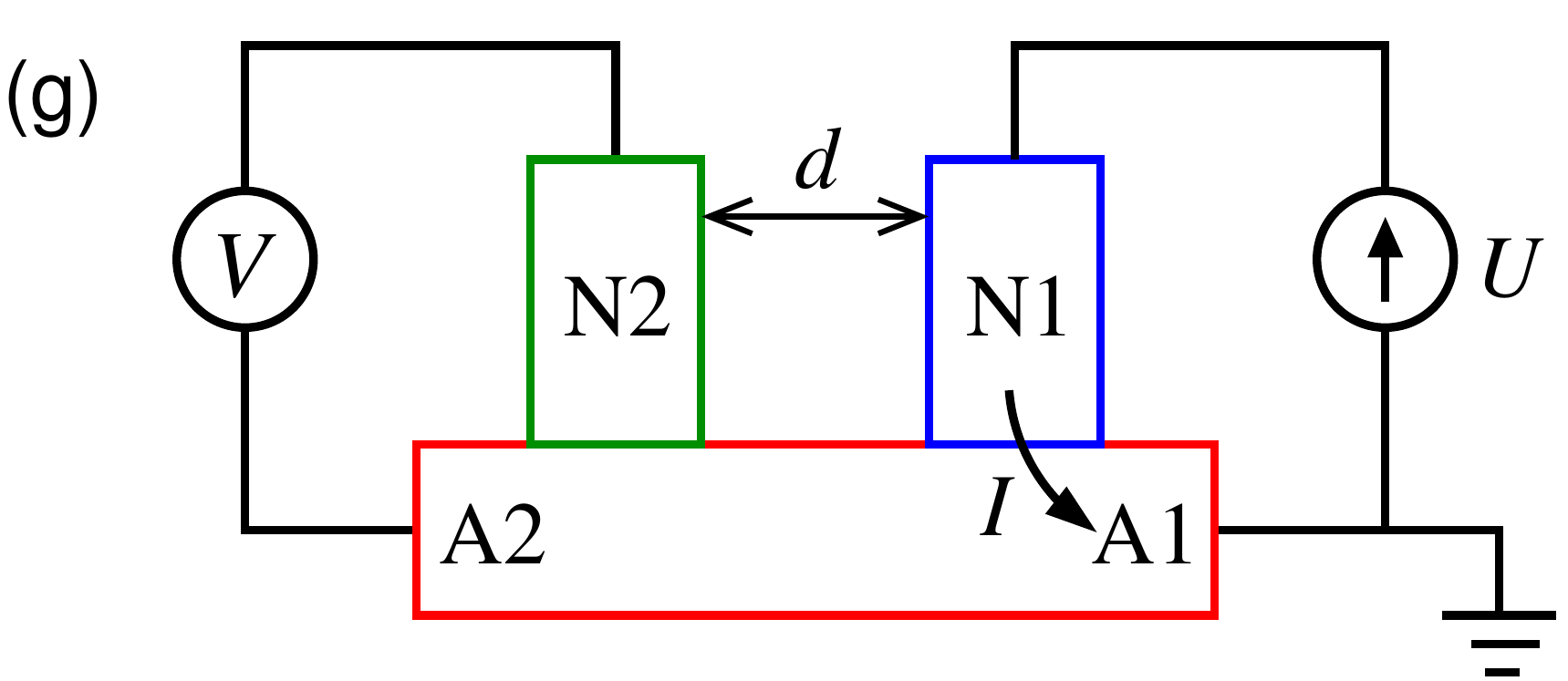}
\caption{Analogy between the AR in N-SC and SAR in N-MI. (a)-(c) AR mechanism: An electron, with energy below the SC gap, incident on the 
interface from the metal can be either reflected (normal reflection), or transmitted via a two-electron process (AR).  The electron enters the SC to form a 
higher energy intermediate state, it then attracts a second electron from the metal to form a Cooper pair, leading to the creation and reflection of a hole 
with opposite spin.  (d)-(f): Analogous SAR mechanism. An electron can be reflected without spin exchange, or it can undergo 
SAR through a second order process. An intermediate doubly occupied state is formed, then an electron of opposite spin is reflected back to the metal. 
Note that the resulting net effect of AR is the transfer of two electrons into the SC, whereas that of the SAR is the spin increase in the MI by $\Delta S^z = 1$. (g) Experimental setup proposed to detect crossed SAR. A1, A2: antiferromagnets; N1, N2: metallic leads; $d$: coherence length; $U$: applied voltage; $I$: flowing current; $V$: measured voltage.}
\label {cartoon}
\end{figure}

Figure~\ref{cartoon} shows an intuitive picture of the proposed SAR 
and a comparison with the AR phenomenon. In the latter, an electron, with energy 
below the SC gap, incident on the interface from the metal side
can be either reflected (normal reflection), 
or transmitted via a two-electron process (AR).  
The electron enters the SC in a higher energy intermediate state, 
then attracts a second electron from the metal to form a Cooper pair, 
eventually leading to the creation 
and reflection of a hole with opposite spin. In the SAR case, 
an electron incident on the interface can either be reflected 
without spin flip, or it undergoes SAR through a second order process. 
An intermediate doubly occupied state is formed, and then an electron of 
opposite spin is reflected back to the metal (with a probability that
depends on parameters like $\overbar U$). 
Note that the net result of AR is the transfer of two electrons 
into the SC, whereas that of the SAR, is the spin increase 
in the MI by $\Delta S^z = 1$.

A few comments are in order.~{\it(1)} In Refs.~\onlinecite{Safi} and~\onlinecite{Daley}, 
the traditional AR is predicted for 1D spinless fermions 
with NN attraction $V$. Using the Jordan-Wigner transformation, 
this model can be mapped into a spin-$\frac{1}{2}$ Heisenberg model. 
The non-interacting chain is mapped into an $XY$ spin chain, whereas the 
superconductor is mapped into a ferromagnetic $XXZ$ chain. Charge AR is 
thus translated into a spin reflection caused by the NN attraction between parallel spins. 
This provides another model in which SAR can be observed, 
although the mechanism of the reflection is different from that of focus 
in this work. {\it (2)} Reference~\onlinecite {Bobkova} proposes 
an interesting $Q$-reflection at N-AFM interfaces. A quasiparticle 
with momentum $\mathbf {k}$ is reflected to 
one with $\mathbf {k} + \mathbf {Q}$ where $\mathbf {Q}$ is the 
ordering vector of the AFM. The scattering phases of spin-up and 
spin-down electrons differ by $\pi$. However, the SAR with spin flip 
is not found mainly because of the mean-field treatment of the AFM. 
{\it (3)} The SAR effect proposed here can also occur in a three-terminal N-MI-N 
geometry (normal-Mott-normal), by analogy with the  N-SC-N 
(normal-SC-normal) investigations already mentioned~\cite{Kleine,CAR}. 
If the width of the SC region is smaller or comparable to the coherence length $\xi_S$, 
the AR induced hole can be generated in the second metal leading to a nonlocal 
charge transfer. In the SAR case, a spin-up electron incident over 
a thin Mott insulator can transmute into a spin-down electron 
on the other side. These two spins will be entangled, with possible 
applications in quantum computing, as in the AR case~\cite{Herrmann,He}. 
{\it (4)} Another area of potential value 
of the SAR effect is in spectroscopy 
where the symmetry of the superconducting state has been studied 
in, e.g., iron-based~\cite{SP} and heavy-fermion~\cite{Laura} superconductors 
via the canonical AR. Exotic Mott states such as $d$-wave
insulators~\cite{YTK} could be analyzed by this procedure. 
In general, any realization currently 
known of the standard AR will admit a translation into the SAR language.

{\it Experimental predictions}.---In recent magnetoconductance experiments on bilayer films of copper (Cu), a normal metal, and copper monoxide, an antiferromagnet, anomalies were observed when compared with Cu grown on a band insulator~\cite{Munakata11}. A proximity effect of antiferromagnetism inside the Cu layer was invoked to explain the results. This effect was theoretically observed before in a real-space dynamical mean-field theory computational study~\cite{Snoek08}, but a simple explanation was not provided. In the context of the SAR effect proposed here, the existence of an ``antiferromagnetic proximity effect'' is natural since the canonical AR is the basis to understand the standard proximity effect in N-SC interfaces~\cite{Pannetier00}. The mapping between the $U>0$ and $U<0$ Hubbard model provides a simple explanation of the results reported in Ref.~\onlinecite{Munakata11}, and it is a concrete prediction of our effort: antiferromagnetic proximity effects into normal metals should be as ubiquitous as in the case of superconductors. Our computational observation that intermediate $|U|$ is more optimal than large $|U|$ to observe the SAR lead us to believe that spin-density-wave (weak coupling) AFM, such as in the recently much investigated iron superconductors, are better than local-spin (strong coupling) AFM materials to test this prediction. 

An even more exotic prediction also emerges from the $U<0$ to $U>0$ mapping. In superconductors the CAR, involving a N-SC-N interface, has attracted considerable attention in quantum entanglement~\cite{Recher01}. The prediction is that a spin-up electron incident on the SC from the first lead may induce a Cooper pair by borrowing a spin-down electron from the second lead, as long as the superconducting width is comparable or smaller than the coherence length. In the canonical CAR a hole current and an associated voltage are produced in the second lead. This voltage is nonlocal in the sense of being in a region without a drive current~\cite{CAR}. Our prediction is that in a similar geometry, the injection of spin-up electrons in an AFM should produce an observable current and voltage of spin-down electrons in the second lead, defining a ``crossed spin Andreev reflection'' (CSAR). The four-terminal geometry employed in Ref.~\onlinecite{Kleine}, replacing the SC by an AFM as in Fig.~\ref{cartoon}~(g), provides a suitable setup to test our prediction.

{\it Conclusion}.---In this Letter, we have predicted a novel 
spin analogue of Andreev reflection in metal-Mott insulator 
heterostructures. An electron incident on the interface from 
the metal can undergo a spin flip upon reflection, thus creating 
a spin-1 excitation in the Mott insulator. This effect was verified 
in 1D models using tDMRG; however, the
proposed mechanism is valid in any dimension. 
Note also that the use of the one-spin-species electron-hole transformation in the Hubbard model merely provides a rapid path to the SAR notion, except for the case of a bipartite lattice where the transformation is exact. However, by mere continuity we believe that models that describe real materials with a density of states close to the particle-hole symmetric point should still display the SAR effect. On the other hand, if the deviation from particle-hole symmetry is large, for example as the doping away from half-filling increases substantially, then the SAR prediction must be revisited.
Also, because of the similarity with the Andreev reflection, this 
effect can be studied in equivalent experimental setups. 
An interesting example is using the spin reflection proposed 
here in order to generate nonlocal entanglement 
as in the crossed Andreev reflection phenomenon.
   
\begin{acknowledgments}
K.A.~thanks G. B. Martins, C. D. Batista, and A. Rahmani for insightful discussions. J.R.~acknowledges fruitful conversations with G. Baskaran. This work was supported by the Center for Nanophase Materials Sciences, sponsored by the U.S.~Department of Energy. K.A., J.R., and G.A.~acknowledge support from the DOE early career research program. E.D.~is supported in part by the U.S.~Department of Energy, Office of Basic Energy Sciences, Materials Science and Engineering Division. J.R.~also acknowledges support by the Simons Foundation (Many Electron Collaboration). Research at Perimeter Institute is supported by the Government of Canada through Industry Canada and by the Province of Ontario through the Ministry of Research and Innovation.

K.A.~and J.R.~contributed equally to this work.
\end{acknowledgments}


\begin{thebibliography}{99}
\bibitem{notebec}
Indeed, for $|U|/t_h < 1$ we observe the formation of long pairs with poor coherence and no evidence for AR. For $|U|/t_h\gg 1$, the on-site pairs (doubly-occupied sites) behave as bosons forming a condensate, suppressing again AR. Therefore, AR is only detected at intermediate coupling.
\end{thebibliography}
\end{document}


\title{
Supplemental Material for\\
Spin Andreev-like Reflection in Metal-Mott Insulator Heterostructures
}

\author{K. A. Al-Hassanieh}
\thanks{These authors contributed equally to this work.}
\affiliation{Center for Nanophase Materials Sciences, Oak Ridge National Laboratory, Oak Ridge, Tennessee 37831, USA}
\author{Juli\'an Rinc\'on}
\thanks{These authors contributed equally to this work.}
\affiliation{Center for Nanophase Materials Sciences, Oak Ridge National Laboratory, Oak Ridge, Tennessee 37831, USA}
\affiliation{Perimeter Institute for Theoretical Physics, Waterloo, Ontario N2L 2Y5, Canada}

\author{G. Alvarez}
\affiliation{Center for Nanophase Materials Sciences, Oak Ridge National Laboratory, Oak Ridge, Tennessee 37831, USA}
\affiliation{Computer Science \& Mathematics Division, Oak Ridge National Laboratory, Oak Ridge, Tennessee 37831, USA}

\author{E. Dagotto}
\affiliation{Materials Science and Technology Division, Oak Ridge National Laboratory, Oak Ridge, Tennessee 37831, USA}
\affiliation{Department of Physics and Astronomy, The University of Tennessee, Knoxville, Tennessee 37996, USA}

\date{\today}

\maketitle



\section{Numerical Tecnique}
The ground state $|\Psi_0\rangle$ is calculated using static DMRG~\cite{White,dmrg}.~Then, a hole Gaussian wave packet with spin $\sigma$ and mean momentum $k_0$ is created in the lead by applying the operator
\begin{equation}
\begin{aligned}
h^\dagger _\sigma (k_0) &= \sum_j {\alpha_j e^{i k_0 j} c_{j\sigma}} \\ &= 
A\sum_j {e^{-(j-j_0)^2/2\sigma_x^2} e^{i k_0 j} c_{j\sigma}}
\end{aligned}
\end{equation}
to $|\Psi_0\rangle$~\cite{Al-Hassanieh13}. The coefficients $\alpha_j$ 
are chosen to generate a Gaussian wave packet centered at $j_0$ with 
width $\sigma_x$. $A$ is a normalization constant. In momentum space, the wave packet is centered at $k_0$ with 
width $\sigma_k = 1/\sigma_x$.  We set $k_0 = k_F - 4\sigma_k$ and $\sigma_k = 0.05$, where $k_F$ is the Fermi momentum of the lead.

The excited state $$|\Psi_e\rangle = h^\dagger_\uparrow(k_0)|\Psi_0\rangle$$ is time evolved under $H$: $|\Psi(t)\rangle = e^{-iHt}|\Psi_0\rangle$. 
We follow the time evolution of the charge and spin densities, $$n_i(t) = n_{i\uparrow}(t) + n_{i\downarrow}(t)$$ and 
$$s_i(t) =\frac{1}{2}\left[n_{i\uparrow}(t) - n_{i\downarrow}(t)\right],$$ respectively with 
$$n_{i\sigma}(t) = \langle\Psi(t)|n_{i\sigma}|\Psi(t)\rangle - \langle\Psi_0|n_{i\sigma}|\Psi_0\rangle,$$
so that, in the results shown, the ground state expectation values are subtracted from the time-dependent measurements. The algorithm used in the time-dependent 
part of the DMRG is based on the Krylov space decomposition~\cite{Schmitteckert04,Manmana05,Schollwoeck06,Boulat08,Alvarez2009,Alvarez2011}. We use $350$ DMRG states in the time evolution calculations, keeping the relative error in the wave 
vector norm of the order of $10^{-3}$ and smaller. The calculations are done for a fixed number of electrons in the system, and $\mu$ is tuned so that the density is uniform. We study systems with $L_I = L_L = 30$ sites.  

\begin{figure}[!b]
\centering
\includegraphics*[width=\columnwidth]{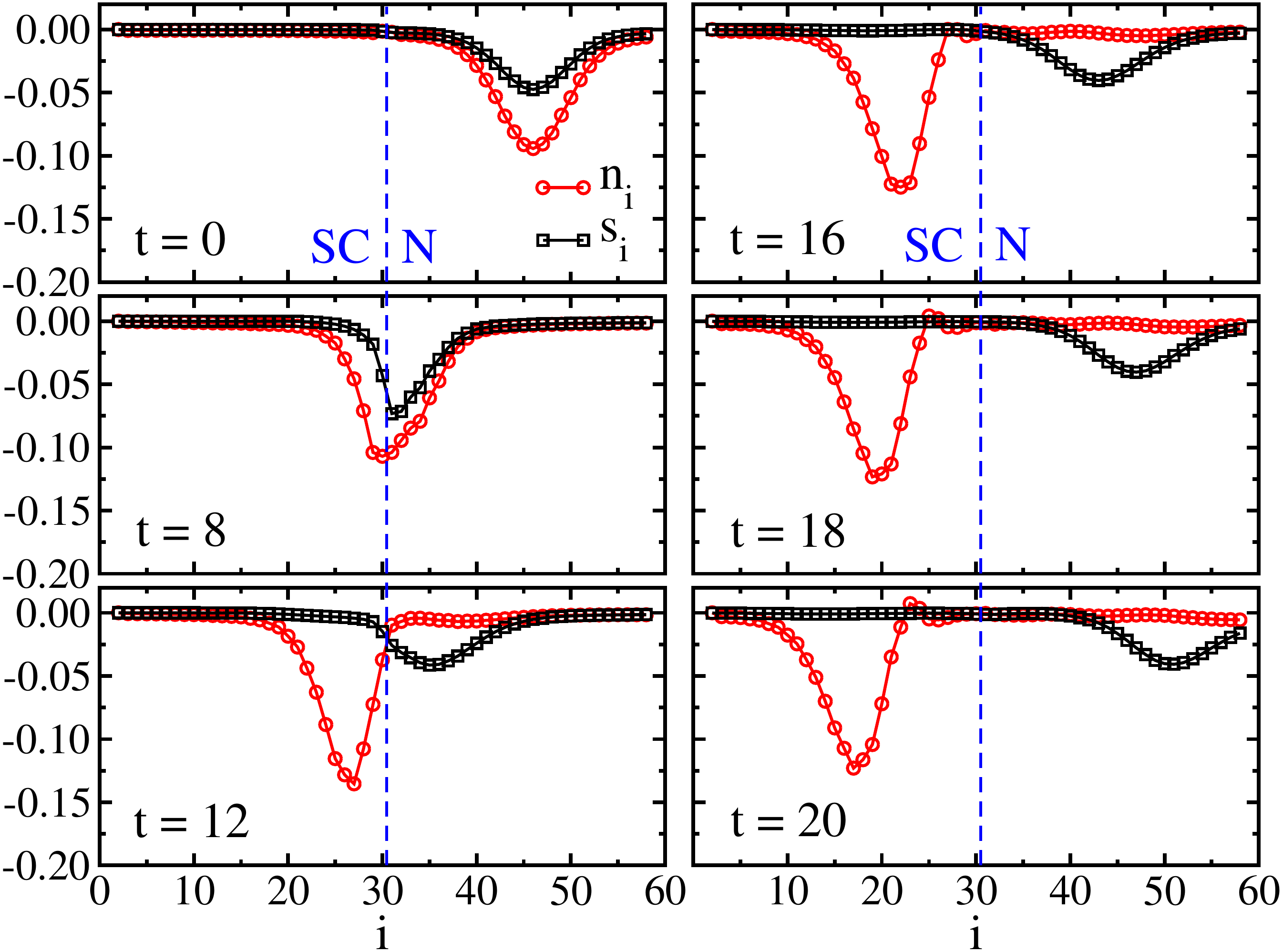}
\caption{Wave packet propagation for $U = -4t_h$ at half filling. $n_i$ and $s_i$ 
are shown at different times. At $t = 0$, the wave packet, with 
charge and spin, is created in the non-interacting lead. It reaches the interface around $t = 8$. 
The charge is transmitted without any reflection, but the spin, on the other hand, 
is reflected completely due to the spin gap.}
\label{Un4Half}
\end{figure}

\section{Half-filled Case Results}
For $U < 0$, the half-filled case was explored for $\mu = 0$, and $N_\uparrow = N_\downarrow = 30$. Figure~\ref {Un4Half} shows the results for the former. At $t = 0$, the wave packet, with charge and spin, is created in 
the non-interacting lead. The charge is transmitted without 
any reflection. The spin, on the other hand, is reflected completely due to the spin gap,
although a small fraction of the spin tunnels 
into the interacting region before it is reflected. 
In the case of half filling, and for $U < 0$, the LL 
parameter is $K_\rho = 1$ which is the same as in the non-interacting 
case. Thus, the charge is transmitted without any scattering, and perfect 
transmission is expected. Similar results are also obtained for larger $|U|$. 
In this case, the ground state of the Hubbard model consists of alternating 
empty and doubly occupied sites. As the charge enters, 
it moves with an effective hopping $t_{\rm eff} \sim t_h^2/|U|$.